\documentclass[journal]{IEEEtran}

\makeatletter
\def\ps@headings{%
\def\@oddhead{\mbox{}\scriptsize\rightmark \hfil \thepage}%
\def\@ adversarynhead{\scriptsize\thepage \hfil \leftmark\mbox{}}%
\def\@oddfoot{}%
\def\@ adversarynfoot{}}
\makeatother
\ifCLASSINFOpdf
\else

\fi

\usepackage{epsfig}
\usepackage{array}
\newcolumntype{L}[1]{>{\raggedright\let\newline\\\arraybackslash\hspace{0pt}}m{#1}}
\newcolumntype{C}[1]{>{\centering\let\newline\\\arraybackslash\hspace{0pt}}m{#1}}
\newcolumntype{R}[1]{>{\raggedleft\let\newline\\\arraybackslash\hspace{0pt}}m{#1}}
\usepackage{amsmath}
\usepackage{amsthm} 
\usepackage{mathrsfs}
\newcommand{\bc}{\begin{center}}
\newcommand{\ec}{\end{center}}
\newcommand{\be}{\begin{equation}}
\newcommand{\ee}{\end{equation}}
\usepackage{flexisym}
\usepackage{algorithmicx}
\usepackage{algorithm}

\usepackage{wasysym}%
\usepackage{soul}
\usepackage{algpseudocode}

\usepackage{amsmath}

\newcommand{\bnu}{\begin{enumerate}}
\newcommand{\enu}{\end{enumerate}}
\usepackage{cite}
\usepackage{url}
\usepackage{breqn}
\usepackage{lipsum}
\usepackage{mathtools}
\usepackage{amsmath}
\usepackage{caption}
\usepackage{subcaption}
\usepackage{soul}
\usepackage{blindtext, graphicx}
\usepackage{cuted}
\usepackage{color}

\usepackage{amsfonts}
\usepackage{lipsum}
\usepackage{cuted}

\begin{document}
\pagestyle{empty}
\title{Dynamic RIS-Assisted THz Quantum Networks: Joint Optimization of Entanglement Generation and Fidelity under Channel Impairments}
\author{Shakil Ahmed,~\IEEEmembership{Member,~IEEE}
\vspace*{-1.05 cm}
\thanks{ Shakil Ahmed is with the Department of Electrical and Computer Engineering, Iowa State University, Ames, Iowa, USA. (email: shakil@iastate.edu).
}}

\maketitle
 
 \thispagestyle{empty}
 
\IEEEpeerreviewmaketitle

\begin{abstract}
Quantum networks (QNs) supported by terahertz (THz) wireless links present a transformative alternative to fiber-based infrastructures, particularly in mobile and infrastructure-scarce environments. However, signal attenuation, molecular absorption, and severe propagation losses in THz channels pose significant challenges to reliable quantum state transmission and entanglement distribution.
To overcome these limitations, we propose a dynamic reconfigurable intelligent surface (RIS)-assisted wireless QN architecture that leverages adaptive RIS elements capable of switching between active and passive modes based on the incident signal-to-noise ratio (SNR). These dynamic RIS elements enhance beamforming control over amplitude and phase, enabling robust redirection and compensation for THz-specific impairments.
We develop a detailed analytical model that incorporates key physical layer phenomena in THz quantum links, including path loss, fading, thermal noise, and alignment variations. A secure optimization framework is formulated to jointly determine RIS placement and entanglement generation rate (EGR) allocation, while satisfying fidelity, security, and fairness constraints under diverse quality of service (QoS) demands. The model also includes an exploration of side-channel vulnerabilities arising from dynamic RIS switching patterns.
Simulation results demonstrate that the proposed architecture yields up to 87\% fidelity enhancement and 65\% fairness improvement compared to static RIS baselines, while maintaining robustness under realistic THz channel conditions. These results underscore the promise of dynamic RIS technology in enabling scalable and adaptive quantum communications over wireless THz links.
\end{abstract}

\section{Introduction}
\IEEEPARstart{Q}{uantum} networks (QNs) facilitate the distribution of entangled quantum states, or qubits, across distant nodes~\cite{chehimi2022physics, ahmed2025osi}, thereby enabling transformative applications such as distributed quantum computing~\cite{cacciapuoti2019quantum}, secure communication via quantum key distribution (QKD)~\cite{chehimi2023quantumMetaverse, chehimi2023roadmap, chehimi2024fedqlstm, ahmed2025quantum, chehimi2024foundations}, and quantum-enhanced sensing~\cite{degen2017quantum}. Unlike classical networks, QNs are constrained by the fragile nature of qubits, limited coherence times, and probabilistic entanglement generation mechanisms~\cite{pirandola2021limits}.
Traditional implementations often rely on optical fiber links for point-to-point quantum communication. However, fixed fiber infrastructure is unsuitable in many emerging deployment scenarios such as mobile platforms, disaster recovery zones, or rural areas with limited connectivity. In such cases, \textit{terahertz (THz) wireless quantum links} offer a promising alternative, offering higher bandwidth, reduced beam divergence, and compatibility with on-chip quantum emitters~\cite{akyildiz2022teranets, sarieddeen2021overview}. Nevertheless, the THz band also introduces challenges, including severe path loss, molecular absorption, and sensitivity to alignment and blockages~\cite{tekbiyik2022terahertz, chaccour2022seven}.

To address these impairments, reconfigurable intelligent surfaces (RISs) have emerged as an enabling technology to dynamically manipulate the wireless channel by altering the phase and amplitude of incident signals~\cite{wu2019intelligent, basar2019wireless, ahmed2024optimizing}. However, existing RIS implementations in QNs assume passive reflection with fixed gain, which limits their adaptability in rapidly changing THz environments. 
In this work, we propose a new class of \textit{dynamic RIS-assisted wireless quantum networks}, where each RIS element can toggle between active and passive modes based on the incoming signal-to-noise ratio (SNR). Inspired by recent advances in RIS circuit design~\cite{zhao2020intelligent, tang2020wireless}, these elements integrate varactor- and tunnel-diode-based structures to support low-power RF amplification and phase control. This hybrid capability allows each element to selectively enhance weak quantum signals while minimizing noise, thereby improving the fidelity of entanglement distribution.

Furthermore, the integration of dynamic RIS into THz-based quantum networks introduces emerging side-channel vulnerabilities that must be carefully addressed to ensure secure quantum communication. Dynamic RIS elements, which actively switch between passive and active modes based on real-time SNR measurements, may unintentionally emit electromagnetic (EM) signatures during switching operations. These signatures can correlate with the internal quantum state distribution processes and may be exploited by adversaries to extract sensitive information. Recent work in side-channel attacks against quantum systems, such as QKD protocols, has demonstrated that machine learning techniques can analyze far-field emissions from single-photon detectors to recover secret keys with high accuracy~\cite{ghosh2024side}. This raises similar concerns in RIS-assisted quantum links, where mode-switching behavior could be profiled by attackers through EM radiation analysis.
Moreover, despite the highly directional nature of THz communication, recent findings show that side-lobe leakage and imperfect beamforming can be leveraged to intercept data even from well-aligned narrow-beam links~\cite{pirandola2021limits}. These vulnerabilities underscore the need to incorporate realistic wireless propagation effects, including molecular absorption, path loss, and scattering, into RIS-assisted quantum channel models. To address these issues, we propose a holistic framework that integrates side-channel security modeling with traditional entanglement fidelity optimization. Our approach ensures both high communication fidelity and resilience against leakage.
In addition, we explore mitigation techniques such as randomized dynamic RIS activation sequences, shielding of RIS control units, and the implementation of secure-by-design QKD protocols like measurement-device-independent QKD (MDI-QKD)~\cite{lo2012mdi}. These strategies aim to obscure any deterministic patterns in RIS element switching and eliminate dependencies between quantum state transmission and observable RIS behavior. By doing so, the proposed system enhances the security posture of THz quantum networks, while preserving performance gains in fidelity and fairness under resource-constrained conditions.
This paper addresses the following research questions:
\begin{enumerate}
    \item \textit{How can we establish and maintain a virtual LoS connection between distant nodes communicating over a wireless quantum channel in a multi-user QN, while adapting to environmental impairments such as fading, misalignment, and path loss?}
    \item \textit{How can we efficiently allocate entangled resources from a constrained QBS to maximize fairness and end-to-end (E2E) fidelity across users, under heterogeneous quantum application QoS requirements?}
\end{enumerate}

\subsection{Prior Works}
QNs are envisioned to facilitate secure communication and distributed quantum computing by enabling the sharing of entangled qubits among distant users~\cite{chehimi2022physics}. Traditional implementations have predominantly relied on fiber-based links, which, while offering high fidelity, are constrained by infrastructure limitations and lack flexibility in dynamic environments~\cite{pirandola2021limits}. To overcome these challenges, wireless quantum communication, particularly in the THz band, has emerged as a promising alternative due to its potential for high data rates and reduced beam divergence~\cite{akyildiz2022teranets}.
However, THz wireless channels introduce unique challenges, including severe path loss, molecular absorption, and sensitivity to alignment and blockages~\cite{sarieddeen2021overview}. To mitigate these issues, RISs have been proposed as a means to dynamically manipulate the wireless environment, enhancing signal propagation and establishing virtual LoS links~\cite{wu2019intelligent}. While initial studies focused on static or passive RIS configurations, recent advancements have introduced dynamic RIS architectures capable of switching between active and passive modes, thereby adapting to real-time channel conditions~\cite{zhao2020intelligent}.

In the context of quantum communication, integrating RISs into QNs has been explored to enhance key distribution and entanglement distribution efficiency. For instance, the use of RISs in multi-user QKD systems has been investigated to address challenges related to decoherence and signal blockage~\cite{kisseleff2023trusted}. These studies highlight the potential of RISs to improve the robustness and scalability of QNs.
Moreover, the interplay between RISs and THz communication has been a subject of recent research. Surveys have analyzed the role of RISs in THz systems, emphasizing their capacity to address propagation challenges and enhance system performance~\cite{ahmed2023survey}. Experimental validations, such as the 220 GHz RIS-aided multi-user THz communication system, have demonstrated the practical feasibility and benefits of integrating RISs into THz networks~\cite{hou2025prototype}.

Despite these advancements, several gaps remain in the literature. Existing works often overlook the joint optimization of RIS placement and quantum resource allocation, particularly in multi-user scenarios with heterogeneous QoS requirements. Additionally, the security implications of dynamic RIS operations, including potential side-channel vulnerabilities, have not been thoroughly investigated. Addressing these challenges necessitates a comprehensive framework that integrates dynamic RIS capabilities, realistic THz channel modeling, and security considerations to optimize entanglement distribution and maintain high fidelity across users.
This paper aims to fill these gaps by proposing a dynamic RIS-assisted wireless quantum network framework that jointly optimizes RIS configuration and entanglement resource allocation. The proposed model accounts for practical THz channel impairments and incorporates security measures to mitigate side-channel threats, thereby enhancing the reliability and scalability of future quantum communication systems.

\subsection{Contributions}
This paper presents a novel architecture for a dynamic RIS-assisted THz-enabled wireless QN, designed to support multi-user entanglement distribution with fairness and robustness under heterogeneous quantum application requirements. Unlike conventional optical or static RIS-based approaches, our framework jointly considers quantum noise, signal fading, misalignment, and the adaptivity of RIS hardware to environmental conditions. The primary contributions are summarized as follows:

\begin{itemize}
    \item We propose the first wireless QN framework that integrates dynamic RIS elements—capable of switching between active and passive reflection states—into THz-based quantum networks to support virtual LoS formation and mitigate link degradation by adjusting phase/amplitude adaptively in response to signal strength and noise conditions.

    \item We develop a comprehensive wireless quantum channel model that accounts for THz-specific impairments including molecular absorption, severe path loss, small-scale fading, and misalignment. Our model introduces quantum-aware metrics such as fidelity and EGR under realistic propagation conditions and supports closed-form derivation of success probability and fidelity bounds for entangled qubits.

    \item We incorporate security-aware considerations by identifying and analyzing side-channel vulnerabilities arising from dynamic RIS reconfiguration patterns, which could potentially leak quantum state behavior to eavesdroppers. We propose safeguards to maintain quantum state confidentiality and robustness in RIS-aided THz systems.

    \item We formulate a joint optimization problem to determine the optimal RIS placement and EGR allocation while satisfying fidelity and rate constraints. The model enforces fairness through weighted Jain’s index objectives and accommodates dynamic RIS behavior. A simulated annealing-based metaheuristic algorithm is developed to solve the non-convex problem efficiently with scalability to multi-user scenarios.

\end{itemize}

Simulation results demonstrate that our proposed framework significantly outperforms classical, fidelity-agnostic allocation baselines. While traditional methods fail to meet minimum fidelity for more than 84\% of users, our framework ensures successful transmission for all users with fidelity above the required thresholds. Furthermore, we achieve up to a 63\% improvement in fairness between users compared to rate-maximization schemes, and show that environmental conditions (e.g., transition from sunny to rainy) can cause up to 49\% degradation in E2E sum rate.

\textbf{Paper Organization} Section~\ref{sec:system_model} presents the system model, including QN components and wireless channel modeling. Section~\ref{sec:loss_noise_analysis} discusses the impact of quantum noise and fading. Section~\ref{sec:optimization} formulates the joint RIS-EGR optimization framework. Section~\ref{sec:simulations} provides simulation results and analysis. Section~\ref{sec:conclusion} concludes the paper and outlines future research directions.

\section{System Model} \label{sec:system_model}
\begin{figure}[ht]
\centering
\includegraphics[width=0.95\linewidth]{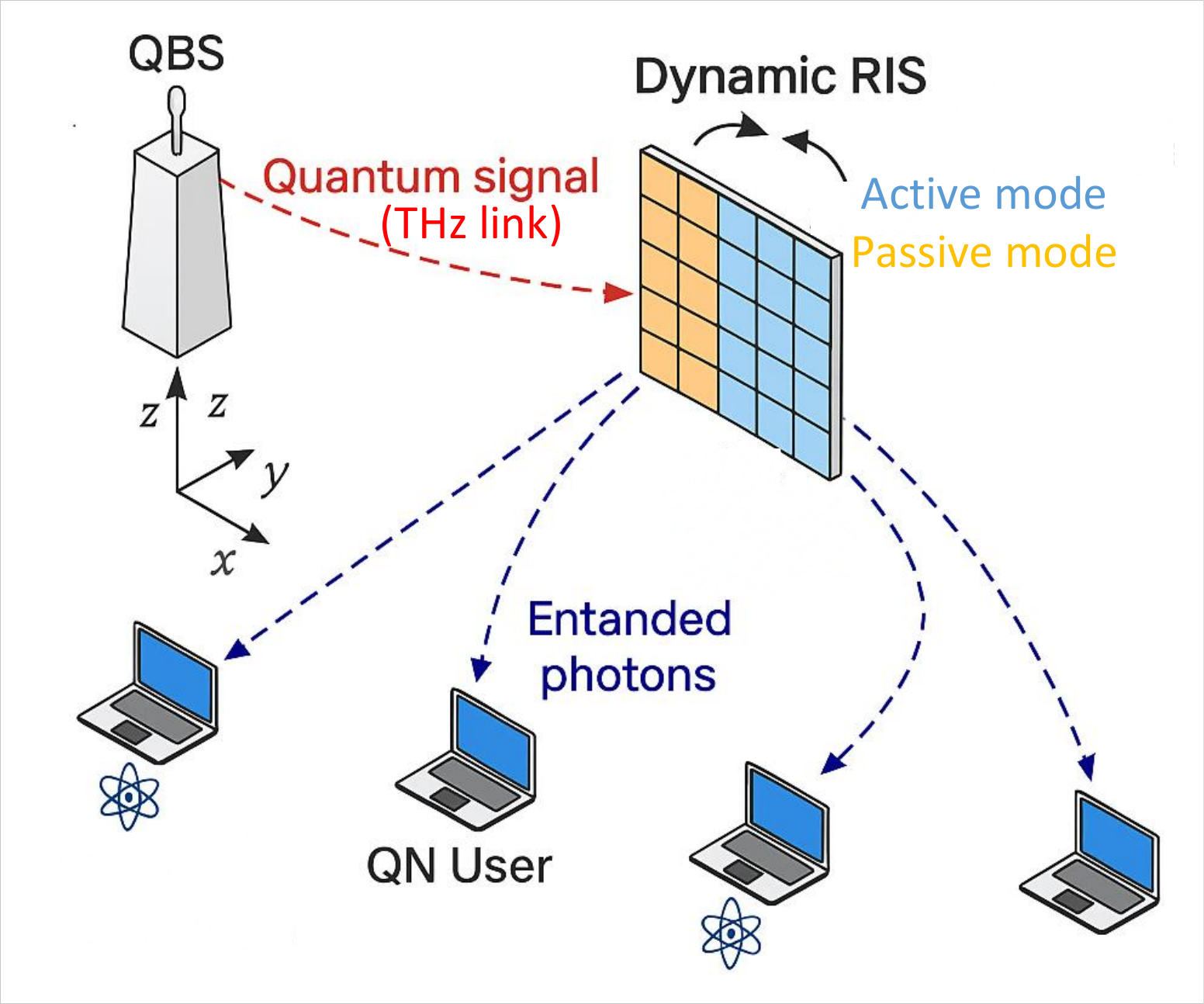}
\caption{System model.}
\label{fig:SM}
\end{figure}

\subsection{System Model with THz Wireless and Dynamic RIS}
As shown in Fig.~\ref{fig:SM}, we consider a terrestrial QN architecture, where a central Quantum Base Station (QBS) communicates with a set $\mathcal{N}$ of $N$ end users via THz-band wireless quantum channels. Each user may participate in distinct quantum applications—such as QKD, distributed sensing, or entanglement-assisted communication—with minimum QoS requirements defined by E2E entanglement fidelity and rate thresholds. Due to terrain irregularities, obstructions, or mobility, these users are assumed to reside in the non-LoS zone of the QBS. To overcome these LoS limitations, a dynamic RIS is deployed to assist entangled qubit transmission by creating virtual LoS paths using programmable beam reflection and focusing.
The QBS, positioned at $\mathbf{l}_s = (x_s, y_s, H_s)$, generates discrete-variable entangled qubit pairs, which are distributed via the RIS, placed at location $\mathbf{l}_r = (x_r, y_r, H_r)$, to user $i \in \mathcal{N}$ located at $\mathbf{l}_i = (x_i, y_i, H_i)$. The RIS plays a dual role: (i) it spatially redirects quantum signals to users, and (ii) its elements dynamically switch between passive (energy-conserving) and active (signal-amplifying) modes, depending on the instantaneous SNR, modeled as a binary function using an SNR threshold~\cite{huang2022reconfigurable, wu2019intelligent}. These switching decisions help optimize signal quality under THz-specific impairments such as molecular absorption, fading, and beam misalignment, while keeping RIS operation power-efficient.
We define the feasible 3D deployment region $\Omega$ for the RIS as:
\begin{align}
\label{eq:deployment_region}
\Omega = \big\{ (x_r, y_r, H_r) \,\big|\,& x_{\min} \leq x_r \leq x_{\max},\,
y_{\min} \leq y_r \leq y_{\max}, \nonumber \\
& H_{\min} \leq H_r \leq H_{\max} \big\},
\end{align}
where these bounds reflect building constraints, mounting possibilities, and signal coverage criteria.
The total E2E distance for user $i$ is the sum of the QBS-to-RIS and RIS-to-user links:
\begin{align}
d_{s,r} &= \sqrt{(x_r - x_s)^2 + (y_r - y_s)^2 + (H_r - H_s)^2}, \\
d_{r,i} &= \sqrt{(x_i - x_r)^2 + (y_i - y_r)^2 + (H_i - H_r)^2}, \\
d_{\text{E2E},i} &= d_{s,r} + d_{r,i}.
\end{align}

Each RIS sub-surface serves one user and adjusts its reflection matrix $\boldsymbol{\Theta}_i$ or $\boldsymbol{\Phi}_i$ based on passive or active state, respectively. The RIS operates under far-field approximation, simplifying the beamforming control to depend solely on 3D location $\mathbf{l}_r$.
Importantly, the use of dynamically switching RIS elements introduces side-channel vulnerabilities: attackers may infer user positions, RIS control signals, or even partial quantum state characteristics from physical emissions, timing patterns, or switching noise~\cite{xu2023risSecurity}. We account for such risks by enforcing minimum entanglement fidelity constraints and by regulating RIS switching dynamics to limit information leakage, thereby motivating future cryptographic shielding techniques at the physical layer.

\subsection{Analysis of RIS Dynamic Element}
In quantum wireless systems assisted by RIS, each meta-atom or programmable element must precisely manipulate incident quantum photonic states. To support reliable entanglement distribution over THz channels, RIS elements operate in either a passive mode, where only phase reconfiguration occurs, or an active mode, where signal amplification is introduced to compensate for loss and decoherence. We define the quantum-aware reflection operations and switching mechanisms associated with these dynamic RIS elements.

\subsubsection{Passive Mode}
In the passive configuration, each RIS element alters the phase and amplitude of incoming quantum signals without injecting external energy. This operation is critical for preserving photon indistinguishability and phase coherence across user channels.
The corresponding reflection coefficient matrix is denoted as:
\begin{equation}
\boldsymbol{\Theta} = \mathrm{diag}\left(\alpha_1 e^{j\theta_1}, \ldots, \alpha_n e^{j\theta_n}, \ldots, \alpha_N e^{j\theta_N} \right),
\end{equation}
where $\boldsymbol{\Theta} \in \mathbb{C}^{N \times N}$ represents the diagonal matrix across $N$ RIS elements. Each element $n$ is characterized by a reflection amplitude $0 \leq \alpha_n \leq 1$ and a programmable phase shift $\theta_n \in [0, 2\pi)$.
Let $s$ be the complex baseband representation of an incident quantum signal (e.g., photonic qubit or weak coherent state). This operation introduces no excess noise or gain, maintaining entanglement fidelity in low-loss regimes. Then, the passively reflected signal from element $n$ at time $t_r$ is modeled as:
\begin{equation}
s_{t_r}^{n} = s \alpha_n e^{j\theta_n}.
\end{equation}

\subsubsection{Active Mode}
When passive reflection is insufficient due to excessive path loss or alignment error, the RIS element transitions to an active mode. Here, amplification is applied in addition to phase control, enabling better signal quality at the cost of potential decoherence or quantum noise.
The active-mode reflection matrix is expressed as:
\begin{equation}
\boldsymbol{\Phi} = \mathrm{diag}\left(\beta_1 e^{j\phi_1}, \ldots, \beta_n e^{j\phi_n}, \ldots, \beta_N e^{j\phi_N} \right),
\end{equation}
where $\boldsymbol{\Phi} \in \mathbb{C}^{N \times N}$, $\beta_n > 1$ is the gain applied to the reflected signal in active mode, and $\phi_n \in [0, 2\pi)$ is the corresponding phase shift.
The reflected quantum signal for element $n$ at time $t_a$ is thus:
\begin{equation}
s_{t_a}^{n} = s \beta_n e^{j\phi_n}.
\end{equation}

While active mode helps improve reach and reliability, it requires strict synchronization to limit quantum error and to satisfy no-cloning and no-amplification constraints for specific quantum protocols (e.g., QKD or teleportation).

\subsubsection{Mode Switching Mechanism}
RIS elements dynamically determine their mode based on local quantum signal quality, typically monitored via a SNR estimator adapted for low-power, photon-level measurements. A binary control variable $\delta_n$ is defined for element $n$:
\begin{equation}
\delta_n =
\begin{cases}
0, & \text{if } \gamma_n \geq \gamma_{\text{th}} \quad \text{(passive mode)} \\
1, & \text{if } \gamma_n < \gamma_{\text{th}} \quad \text{(active mode)}
\end{cases},
\end{equation}
where $\gamma_n$ denotes the local SNR observed at RIS element $n$, and $\gamma_{\text{th}}$ is a predefined threshold that reflects minimum entanglement fidelity or photon arrival probability.
The overall behavior of the reflected signal at time $t$ is compactly expressed as:
\begin{equation}
s_t^n = s \left[ (1 - \delta_n) \alpha_n e^{j\theta_n} + \delta_n \beta_n e^{j\phi_n} \right],
\end{equation}
where $\delta_n \in \{0,1\}$ enables mode switching in real time. This binary switching supports entanglement fairness and adaptive quantum beamforming under heterogeneous user conditions and link degradations.

\subsection{Channel Model with Dynamic RIS}
We consider a quantum-aware wireless channel model that captures realistic impairments relevant to THz dynamic RIS-assisted communication systems. The model incorporates path loss, small-scale fading, pointing misalignment, and THz-specific absorption. Additionally, we model the dynamic behavior of RIS elements that can switch between passive and active states based on the observed SNR, enabling flexible control of entangled photon steering. This forms the foundation for high-fidelity, fair, and secure quantum communication across multi-user wireless quantum networks~\cite{huang2022reconfigurable, faisal2023surveyTHz, liu2021reconfigurable}.
The E2E quantum channel gain between the QBS and user $i \in \mathcal{N}$ via the RIS is modeled as:
\begin{equation}
h_i(\mathbf{l}_r) = \varsigma_i \cdot h_i^{\text{pl}}(\mathbf{l}_r) \cdot h_i^{\text{f}}(\mathbf{l}_r) \cdot h_i^{\text{p}}(\mathbf{l}_r),
\end{equation}
where  $\varsigma_i$ is the RIS reflection efficiency for user $i$, dependent on the RIS element state;
  $h_i^{\text{pl}}$ models the THz path loss with absorption;
  $h_i^{\text{f}}$ captures small-scale fading;
  $h_i^{\text{p}}$ models alignment error due to beam pointing deviations.

\subsubsection{THz Path Loss}
In the THz band, molecular absorption and distance attenuation dominate. We adopt the path loss model:
\begin{equation}
h_i^{\text{pl}}(\mathbf{l}_r) = \left( \frac{\lambda}{4 \pi d_{\text{E2E},i}(\mathbf{l}_r)} \right)^2 \cdot \exp\left(-\kappa(f) \cdot d_{\text{E2E},i}(\mathbf{l}_r)\right),
\end{equation}
where $\lambda$ is the carrier wavelength, $d_{\text{E2E},i}$ is the QBS-to-user distance via the RIS, and $\kappa(f)$ is the molecular absorption coefficient at frequency $f$~\cite{rappaport2019wireless}.

\subsubsection{Small-Scale Fading}
Due to sparse reflectors and strong directivity at THz frequencies, we model fading using the Rician distribution:
\begin{equation}
h_i^{\text{f}}(\mathbf{l}_r) \sim \text{Rician}(K, \sigma),
\end{equation}
where $K$ is the Rician factor indicating the strength of the line-of-sight (LoS) path relative to the scattered components, and $\sigma^2$ is the average fading power~\cite{ghosh2022alignment}.

\subsubsection{Pointing Misalignment}
THz links suffer from beam misalignment, especially for narrow beams. We use a Gaussian-based pointing loss model:
\begin{equation}
h_i^{\text{p}}(\mathbf{l}_r) = A_0 \cdot \exp\left(-\frac{2\Delta\theta_i^2}{\omega_{\text{eq},i}^2}\right),
\end{equation}
where $\Delta\theta_i$ is the angular deviation from the beam center and $\omega_{\text{eq},i}$ is the equivalent beamwidth of the RIS-user link~\cite{ghosh2022alignment}.

\subsubsection{RIS Reflection Efficiency and Mode Switching}
Each RIS element dynamically switches between active and passive states based on its local SNR. Let $\delta_n \in \{0, 1\}$ denote the state of element $n$:
\begin{equation}
\varsigma_i =
\begin{cases}
\varsigma_{\text{passive}}, & \text{if } \text{SNR}_i \geq \gamma_{\text{th}} \\
\varsigma_{\text{active}}, & \text{if } \text{SNR}_i < \gamma_{\text{th}}
\end{cases},
\end{equation}
where $\gamma_{\text{th}}$ is a predefined SNR threshold, and $\varsigma_{\text{passive}} < 1 < \varsigma_{\text{active}}$. 
\subsection{Side-Channel Leakage}
Dynamic switching patterns introduce time-varying reflection coefficients that may leak side-channel information to a passive adversary. Let $\mathbf{\varsigma}(t) = [\varsigma_1(t), \ldots, \varsigma_N(t)]$ be the RIS reflection vector over time. The mutual information between $\mathbf{\varsigma}(t)$ and an adversary’s observation $\mathbf{Z}$ can be quantified as:
\begin{equation}
\mathcal{L}_{\text{leak}} = I(\mathbf{\varsigma}(t); \mathbf{Z}) > 0,
\end{equation}
indicating potential information leakage if patterns are predictable~\cite{wang2023risattack}. To mitigate this, randomized coding or phase hopping can be introduced such that:
\begin{equation}
\mathcal{L}_{\text{leak}} \approx 0 \quad \text{subject to } \mathbb{E}[\text{Fidelity}] \geq F_{\min},
\end{equation}
where $F_{\min}$ is the minimum acceptable fidelity for secure entanglement. Such techniques enhance stealth and confidentiality without compromising quantum link performance.

\section{Entanglement Generation and Distribution} \label{sec:loss_noise_analysis}
\begin{figure}[]
    \centering
    \includegraphics[width=0.9\linewidth]{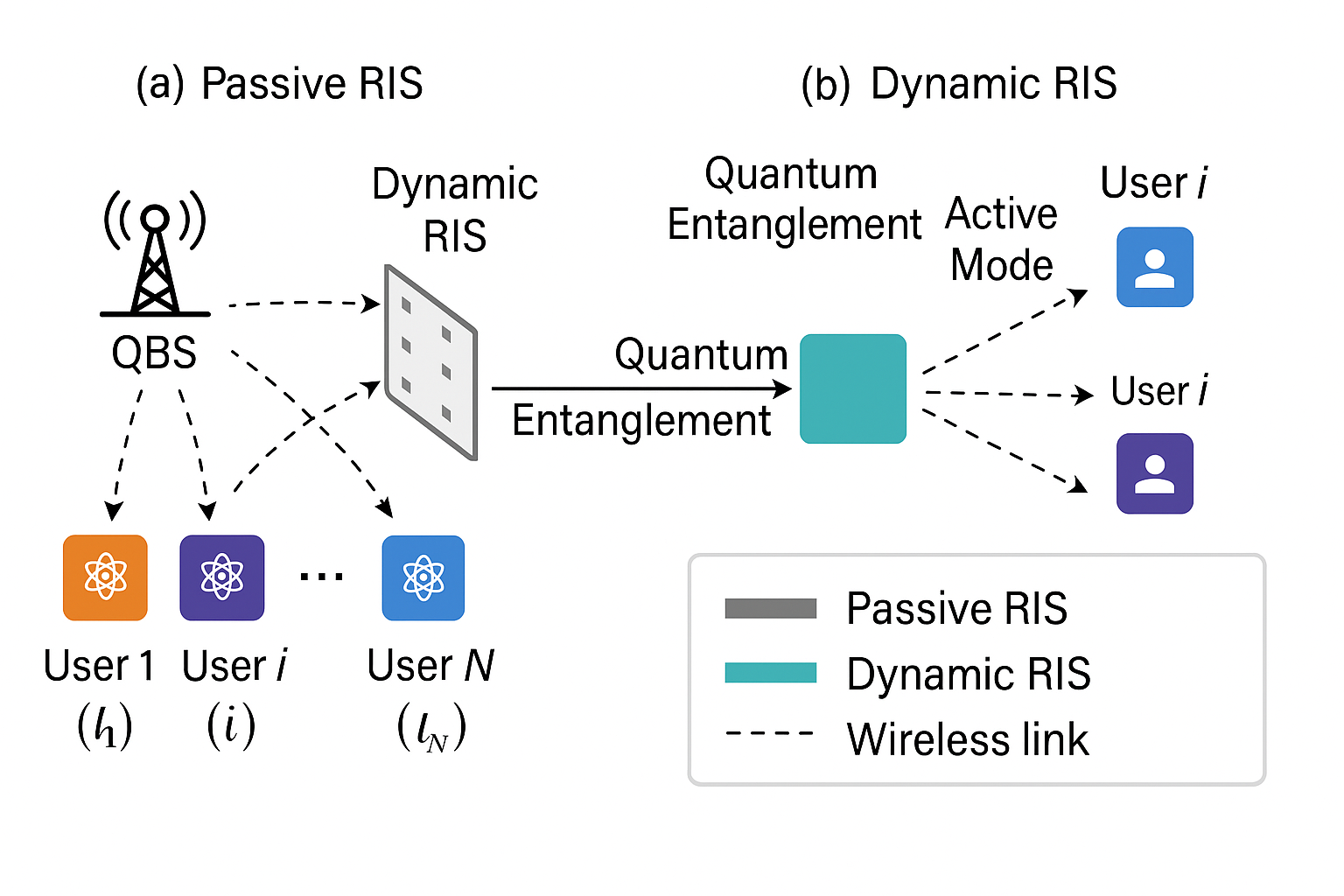}
    \caption{System-level diagram illustrating dynamic RIS-assisted quantum signal routing, where RIS adaptively switches modes based on signal strength to enhance fidelity and entanglement.}
    \label{fig:dynamic_ris_quantum}
\end{figure}

In the proposed THz-enabled star-shaped QN, the central QBS is equipped with an entanglement source capable of generating discrete-variable entangled qubit pairs. Each entangled pair consists of: (i) a matter qubit, which is stored in a local quantum memory at the QBS, and (ii) a flying qubit, which is transmitted as a photonic carrier through a THz-band wireless quantum channel via the dynamic RIS. This architecture supports a maximum generation and buffering rate of $C_{\max}$ entangled pairs per second, constrained by the memory's coherence time and storage capacity.
The matter qubit remains localized in the QBS memory and is subject to decoherence due to interactions with its environment, including thermal noise and memory instability. These imperfections degrade quantum fidelity over time, and the effective storage time is bounded by the quantum memory's coherence time $T_{\text{coh}}$. Importantly, loss events are minimal for matter qubits, but their degradation arises primarily from quantum phase noise and amplitude damping.
On the other hand, the flying qubit must traverse a lossy and noisy THz wireless quantum link to reach its intended user. Upon emission, the flying qubit is directed to the RIS, where it undergoes programmable reflection—either in passive or active mode—depending on the local SNR. This RIS-enabled reflection dynamically optimizes the spatial steering, gain, and phase alignment to mitigate beam divergence, THz molecular absorption, and pointing error.
The total EGR for user $i \in \mathcal{N}$ is thus limited by the probability of successful photon delivery through the end-to-end RIS-assisted THz link. Denote $\eta_i$ as the E2E transmissivity (or photonic success probability), which captures path loss, RIS reflection efficiency, and receiver aperture coupling. It can be expressed as:
\begin{equation}
\eta_i = \varsigma_i \cdot h_i^{\text{pl}}(\mathbf{l}_r) \cdot h_i^{\text{f}}(\mathbf{l}_r) \cdot h_i^{\text{p}}(\mathbf{l}_r),
\end{equation}
where each term is defined in Section~III. The effective EGR is then given by:
\begin{equation}
\text{EGR}_i = \min \left( C_{\max}, \, R_i \cdot \eta_i \right),
\end{equation}
where $R_i$ is the qubit transmission rate allocated to user $i$.
While loss reduces the success probability, noise (e.g., thermal, phase, or RIS-amplification noise) directly degrades the fidelity $F_i$ of the shared entangled state between the QBS and user $i$. We define a fidelity floor $F_{\min}$, below which entangled states are considered unusable for quantum applications such as QKD. The RIS plays a crucial role in preserving $F_i$ by adaptively selecting amplification gain (active mode) or low-noise reflection (passive mode) depending on the signal's propagation conditions. This switching is shown in Fig.~\ref{fig:dynamic_ris_quantum}.
The use of dynamic RIS elements enables quantum link adaptation without invasive amplification or intrusive measurement. Passive mode minimizes added noise, preserving fragile quantum coherence, while active mode helps overcome deep fading or absorption by injecting controlled gain. This dual-mode control supports fair entanglement distribution across heterogeneous users and compensates for path variability in mobile or obstructed environments.
The RIS-assisted architecture enhances quantum entanglement delivery over THz links by intelligently balancing transmission success (loss mitigation) and fidelity preservation (noise control), making it a viable framework for practical multi-user quantum networking in open-air or non-LoS deployments.

\subsection{Losses and EGR Analysis}
In the proposed THz-assisted quantum network, the QBS is responsible for generating and fairly distributing entangled qubits to end users $i \in \mathcal{N}$. Due to the absence of a LoS between the QBS and the users, the entangled photonic qubits are directed via a dynamically controlled RIS, which establishes virtual LoS links through programmable beam reflection and amplification.
At the QBS, each entangled qubit pair consists of a matter qubit stored in quantum memory and a flying qubit (i.e., photonic qubit) sent through the wireless THz channel toward user $i$ via the RIS. Let $R_{\text{in},i}$ denote the initial rate of entangled pair generation dedicated to user $i$, measured in pairs per second. However, due to propagation impairments such as THz path loss, small-scale fading, and pointing misalignment, only a subset of these flying qubits are successfully delivered, leading to a reduced effective EGR.
To quantify this, we define the probability of successful photon delivery from the QBS to user $i$ via the RIS as $P_{\text{succ},i}(\mathbf{l}_r)$, where $\mathbf{l}_r$ is the 3D RIS location. This probability encapsulates the combined effects of: \textit{1) Distance-dependent THz path loss, 2) Small-scale multipath fading,, and 3) Pointing error due to RIS misalignment.}
We define a success event as the case where the total RIS-assisted E2E channel gain $h_i(\mathbf{l}_r)$ exceeds a fidelity-dependent threshold $\zeta_{\text{th}}$, ensuring the quantum state is usable for downstream applications such as QKD.

\textit{Theorem 1:} \textit{The probability of successfully transmitting a flying qubit from the QBS to user $i$ over the RIS-assisted THz quantum channel is given by:}
\begin{align}
& P_{\text{succ},i}(\mathbf{l}_r) = 1 
- \left( \frac{\vartheta_i(\mathbf{l}_r)}{\Gamma(\alpha_i(\mathbf{l}_r)) \Gamma(\beta_i(\mathbf{l}_r))} \right) \nonumber \\
&
\mathcal{G}_{2,4}^{3,1} \left[
\left. \frac{\alpha_i(\mathbf{l}_r) \beta_i(\mathbf{l}_r) \chi_{\text{th}}}
{A_{0,i}(\mathbf{l}_r) h_i^{\text{pl}}(\mathbf{l}_r)} 
\right| 
\begin{array}{l}
1, \quad \vartheta_i(\mathbf{l}_r)+1 \\
\vartheta_i(\mathbf{l}_r), \alpha_i(\mathbf{l}_r), \beta_i(\mathbf{l}_r), 0
\end{array}
\right], \label{eq:succ}
\end{align}
\noindent where $\mathcal{G}^{p,q}_{m,n}[\cdot]$ is the Meijer’s G-function, $\Gamma(\cdot)$ is the Gamma function, and $\chi_{\text{th}} = \frac{\zeta_{\text{th}}}{\eta}$ represents a normalized gain threshold based on quantum detection sensitivity $\eta$.

\textit{Proof:} See Appendix~A. \hfill

\noindent As evident from Theorem~1, the probability of successful delivery depends heavily on environmental factors (e.g., RIS placement, absorption coefficient, and turbulence) and the THz channel characteristics. The Meijer G-function generalizes the distribution of the composite gain and facilitates closed-form tractability.
Finally, the average rate of successfully distributed entangled pairs (measured in Hz) to user $i$—referred to as the effective EGR—is given by:
\begin{equation}
R_{\text{E2E},i}(\mathbf{l}_r, R_{\text{in},i}) = P_{\text{succ},i}(\mathbf{l}_r) \cdot R_{\text{in},i}, \label{eq:e2e}
\end{equation}
where $R_{\text{E2E},i}$ captures the probabilistic throughput of entangled qubits reaching user $i$ with acceptable fidelity, considering RIS location and dynamic beam steering behavior. This metric plays a key role in evaluating fairness and reliability of entanglement distribution across the network.

\subsection{Noise and E2E Entangled State Analysis}
Each entangled qubit pair generated by the QBS and assigned to user $i \in \mathcal{N}$ is described by a Bell-diagonal mixed state~\cite{nielsen2010quantum}, which arises from the interaction of ideal entangled states with environmental noise. The density matrix of such a state is given by:
\begin{equation}
\rho_{\text{BD},i} = \sum_{m,n \in \{0,1\}} \lambda_{mn,i} \Phi_{mn}, \label{eq:bell_state}
\end{equation}
where $\Phi_{mn} = |\Phi_{mn} \rangle \langle \Phi_{mn}|$ and $|\Phi_{mn} \rangle = (\sigma_x^m \sigma_z^n \otimes \mathbb{I}) |\Phi^+ \rangle$, with $|\Phi^+ \rangle = \frac{1}{\sqrt{2}} (|00\rangle + |11\rangle)$ the canonical maximally entangled state. The Pauli operators $\sigma_x$ and $\sigma_z$ introduce bit and phase flips, and the eigenvalues $\lambda_{mn,i}$ satisfy $\sum_{m,n} \lambda_{mn,i} = 1$.
The fidelity of $\rho_{\text{BD},i}$ with respect to $\Phi_{00}$, which serves as the target entangled state for quantum applications (e.g., QKD, teleportation), is:
\begin{equation}
f_{\text{target},i} = \langle \Phi_{00} | \rho_{\text{BD},i} | \Phi_{00} \rangle = \lambda_{00,i}.
\end{equation}

To ensure reliable quantum processing, the distributed E2E entangled state must achieve a fidelity above a specified threshold. However, as each half of the entangled pair undergoes a different physical process—storage for the matter qubit and THz wireless transmission for the flying qubit—distinct noise sources affect their quality. The flying qubit is particularly susceptible to channel impairments such as loss, turbulence, and pointing misalignment, which are modeled as quantum noise channels.

\textit{Remark 1:} The composite noise experienced by the flying qubit transmitted over a THz quantum wireless channel with RIS-assisted redirection can be modeled as a concatenation of phase damping and amplitude damping channels.

The phase damping channel captures decoherence due to phase fluctuations (e.g., from turbulence), and is defined as:
\begin{equation}
\Lambda_{i}^{(1)}(\rho) = (1 - P_{\text{pd},i}(\mathbf{l}_r)) \rho + P_{\text{pd},i}(\mathbf{l}_r) \sigma_z \rho \sigma_z, \label{eq:phase_damping}
\end{equation}
where $P_{\text{pd},i}(\mathbf{l}_r)$ is the phase error probability induced by turbulence, modeled via the Rytov variance as:
\begin{equation}
P_{\text{pd},i}(\mathbf{l}_r) = \text{erf}(\sigma_{R,i}^2(\mathbf{l}_r)) = \frac{2}{\sqrt{\pi}} \int_0^{\sigma_{R,i}^2(\mathbf{l}_r)} e^{-t^2} dt,
\end{equation}
with $\sigma_{R,i}^2(\mathbf{l}_r) = 1.23 C_n^2 k^{7/6} d_{\text{E2E},i}^{11/6}(\mathbf{l}_r)$, where $C_n^2$ is the turbulence coefficient and $k = 2\pi / \lambda$ is the optical wavenumber.
The amplitude damping channel models energy loss due to absorption and scattering, and is represented as:
\begin{equation}
\Lambda_{i}^{(2)}(\rho) = E_{0,i} \rho E_{0,i}^\dagger + E_{1,i} \rho E_{1,i}^\dagger, \label{eq:amp_damp}
\end{equation}
with Kraus operators:
\begin{align}
E_{0,i} &= |0\rangle \langle 0| + \sqrt{1 - P_{\text{ad},i}(\mathbf{l}_r)} |1\rangle \langle 1|, \\
E_{1,i} &= \sqrt{P_{\text{ad},i}(\mathbf{l}_r)} |0\rangle \langle 1|,
\end{align}
where $P_{\text{ad},i}(\mathbf{l}_r) = 1 - P_{\text{succ},i}(\mathbf{l}_r)$ reflects the complement of the success probability derived in Section~\ref{sec:loss_noise_analysis}.
Together, these channels describe the evolution of the flying qubit during RIS-assisted transmission. The final E2E entangled state, denoted $\rho_{\text{E2E},i}(\mathbf{l}_r)$, results from applying these noise maps to the flying qubit of $\rho_{\text{BD},i}$. The fidelity of the resulting bipartite state is given below.

\textit{Proposition 1:} The E2E fidelity between the QBS and user $i$ over a RIS-assisted THz quantum channel is given by:
\begin{align}
f_{\text{E2E},i}(\mathbf{l}_r) &= \frac{1}{4} \left( U_{00,i} + U_{01,i} \right) \nonumber \\
&+ (1 - P_{\text{succ},i}(\mathbf{l}_r)) \left( U_{10,i} + U_{11,i} \right) \nonumber \\
&+ P_{\text{succ},i}(\mathbf{l}_r) \left( U_{00,i} + U_{01,i} \right) \nonumber \\
&+ \frac{1}{2} \sqrt{P_{\text{succ},i}(\mathbf{l}_r)} \left( U_{00,i} - U_{01,i} \right), \label{eq:fidelity}
\end{align}
where the $U_{jk,i}$ terms account for phase decoherence:
\begin{align}
U_{jk,i}(\mathbf{l}_r) &= \left(1 - P_{\text{pd},i}(\mathbf{l}_r) \right) V_{jk,i}(\mathbf{l}_r)  
+ P_{\text{pd},i}(\mathbf{l}_r) V_{j(k \oplus 1),i}(\mathbf{l}_r), \\
V_{jk,i}(\mathbf{l}_r) &= (1 - p_{1,i}) \lambda_{jk,i} + p_{1,i} \lambda_{j(k \oplus 1),i},
\end{align}
and $p_{1,i}$ models memory decoherence on the matter qubit.

\textit{Proof:} See Appendix~B. \hfill

This fidelity formulation integrates quantum decoherence, RIS-assisted channel losses, and THz-specific impairments into a unified measure of entanglement quality. Importantly, it reveals that RIS deployment location $\mathbf{l}_r$ plays a central role in ensuring reliable E2E quantum connectivity by minimizing phase variance and boosting transmission success.

\subsection{Joint RIS and EGR Allocation Optimization Problem} \label{sec:optimization}
The analysis in Section~\ref{sec:loss_noise_analysis} shows that the quality and rate of entangled qubit distribution are critically influenced by wireless quantum channel impairments, RIS placement, and the dynamic configuration of RIS elements. Particularly, the proposed RIS architecture dynamically switches each element between passive and active modes depending on local channel conditions, as defined by the binary variable $\delta_n \in \{0, 1\}$. This switching enhances signal quality by either preserving quantum coherence in passive mode or compensating attenuation through active amplification when necessary. To fully leverage these capabilities, we now formulate a joint optimization problem incorporating RIS location, EGR resource allocation, and adaptive RIS element behavior.

Let $R_{\text{in},i}$ denote the initial EGR assigned by the QBS to user $i \in \mathcal{N}$. Let $\mathbf{l}_r = (x_r, y_r, H_r)$ denote the 3D RIS placement within the deployment region $\Omega$ defined in~\eqref{eq:deployment_region}, and let $\boldsymbol{\delta} = [\delta_1, \delta_2, \dots, \delta_N]$ represent the binary mode selection of each RIS element. The decision variables are thus: (i) $\mathbf{R}_{\text{in}}$: initial EGR vector, (ii) $\mathbf{l}_r$: RIS position vector, and (iii) $\boldsymbol{\delta}$: dynamic mode profile across the RIS array.
The optimization goal is to maximize the weighted sum of effective E2E entanglement generation rates across all users, while ensuring memory capacity limits, entanglement fidelity, and fairness constraints are satisfied. The utility function is captured using the weighted fairness index (WFI)~\cite{jain1984quantitative}:
\begin{equation}
U_{\text{WFI}}(\mathbf{l}_r, \mathbf{R}_{\text{in}}, \boldsymbol{\delta}) = 
\frac{\left( \sum_{i \in \mathcal{N}} w_i R_{\text{E2E},i}(\mathbf{l}_r, R_{\text{in},i}, \boldsymbol{\delta}) \right)^2}
{\sum_{i \in \mathcal{N}} w_i^2 \left(R_{\text{E2E},i}(\mathbf{l}_r, R_{\text{in},i}, \boldsymbol{\delta}) \right)^2}, \label{eq:wfi_delta}
\end{equation}
where the E2E entanglement rate $R_{\text{E2E},i}$ is now explicitly dependent on both the RIS placement and the configuration of dynamic RIS elements.
The joint optimization problem is formulated as follows:
\begin{subequations}
\begin{align} 
\mathcal{P}_1: \quad & \max_{\mathbf{R}_{\text{in}}, \mathbf{l}_r, \boldsymbol{\delta}} \sum_{i \in \mathcal{N}} w_i R_{\text{E2E},i}(\mathbf{l}_r, R_{\text{in},i}, \boldsymbol{\delta}) \label{eq:opt_obj_delta} \\
\text{s.t.} \quad 
& \sum_{i \in \mathcal{N}} R_{\text{in},i} \leq C_{\text{max}}, \label{eq:cons_memory_delta} \\
& R_{\text{E2E},i}(\mathbf{l}_r, R_{\text{in},i}, \boldsymbol{\delta}) \geq R_{\text{min},i}, \quad \forall i \in \mathcal{N}, \label{eq:cons_rate_delta} \\
& f_{\text{E2E},i}(\mathbf{l}_r, \boldsymbol{\delta}) \geq f_{\text{min},i}, \quad \forall i \in \mathcal{N}, \label{eq:cons_fidelity_delta} \\
& U_{\text{WFI}}(\mathbf{l}_r, \mathbf{R}_{\text{in}}, \boldsymbol{\delta}) \geq \delta_{\text{th}}, \label{eq:cons_fairness_delta} \\
& x_{\min} \leq x_r \leq x_{\max}, \label{eq:cons_x_delta} \\
& y_{\min} \leq y_r \leq y_{\max}, \label{eq:cons_y_delta} \\
& H_{\min} \leq H_r \leq H_{\max}, \label{eq:cons_h_delta} \\
& \delta_n \in \{0,1\}, \quad \forall n = 1,\dots,N. \label{eq:cons_delta}
\end{align}
\end{subequations}

In this formulation, the objective~\eqref{eq:opt_obj_delta} maximizes the total weighted E2E entanglement throughput, considering the joint impact of EGR allocation, RIS position, and RIS mode selection. Constraint~\eqref{eq:cons_memory_delta} limits the total entanglement injected into the network to not exceed the QBS memory budget. Constraint~\eqref{eq:cons_rate_delta} guarantees per-user minimum throughput. Constraint~\eqref{eq:cons_fidelity_delta} enforces fidelity requirements based on both the RIS configuration and the propagation loss model. Constraint~\eqref{eq:cons_fairness_delta} enforces a minimum WFI threshold $\delta_{\text{th}}$. Spatial constraints~\eqref{eq:cons_x_delta}–\eqref{eq:cons_h_delta} define the feasible 3D region for RIS placement, and~\eqref{eq:cons_delta} enforces the binary nature of dynamic RIS mode selection.

\subsubsection{Solution Approach}
The optimization problem formulated in (\ref{eq:opt_obj_delta})--(\ref{eq:cons_delta}) is inherently non-convex due to several reasons: (i) the nonlinear and probabilistic nature of the constraints, especially those involving fidelity $f_{\text{E2E},i}(\mathbf{l}_r)$ and success probability $P_{\text{succ},i}(\mathbf{l}_r)$; (ii) the coupling between continuous optimization variables—such as the RIS location vector $\mathbf{l}_r$ and the entanglement generation rates $\{R_{\text{in},i}\}$—and the discrete dynamic reflection states of the RIS elements, denoted by $\boldsymbol{\delta} = \{\delta_n\}_{n=1}^{N}$; and (iii) the presence of complex special functions like Meijer-G, which hinder closed-form derivation or convex reformulation.
To address these challenges, we adopt a metaheuristic approach based on Simulated Annealing (SA). SA is well-suited for solving large-scale, non-convex, and combinatorial problems by allowing stochastic exploration of the solution space, thereby avoiding local optima. Specifically, the algorithm begins with a randomly initialized feasible solution, which includes the RIS placement $\mathbf{l}_r^{(0)}$ and a feasible configuration of RIS element modes $\boldsymbol{\delta}^{(0)}$, selecting either passive or active operation per element according to their local SNRs. The initial entanglement generation rates $\mathbf{R}_{\text{in}}^{(0)}$ are also chosen to respect the memory capacity and QoS constraints.

At each iteration, a new candidate solution is generated by perturbing one or more of the current variables: RIS location, EGR allocation, or RIS element states. The new candidate $(\mathbf{l}_r', \mathbf{R}_{\text{in}}', \boldsymbol{\delta}')$ is evaluated by computing the associated end-to-end rate $R_{\text{E2E},i}$ and fidelity $f_{\text{E2E},i}$ for all users. If all constraints are satisfied and the weighted fairness utility improves, the candidate is accepted. Otherwise, it is accepted with a probability determined by a Boltzmann distribution, $\exp(-\Delta/T)$, where $\Delta$ is the deterioration in objective value and $T$ is the current temperature. The temperature follows a cooling schedule, typically geometric, which gradually reduces the chance of accepting worse solutions.
This annealing process continues until a convergence criterion is met—either a maximum number of iterations or stagnation in objective improvement. The algorithm effectively balances exploration and exploitation, enabling the discovery of high-quality solutions even under the realistic impairments imposed by THz wireless quantum channels and dynamically reconfigurable RIS elements. The detailed computational steps of this SA-based solution procedure are summarized in Algorithm~\ref{alg:algorithm_ris}.

\subsubsection{Proposed Algorithm}
The optimization problem $\mathcal{P}_1$ is non-convex and NP-hard due to its coupling of continuous (e.g., RIS placement and entanglement generation rates) and discrete variables (e.g., RIS element states), as well as nonlinear probabilistic constraints. To efficiently tackle this problem, we adopt a SA-based metaheuristic algorithm. SA is capable of escaping local optima through probabilistic acceptance of worse solutions during the early high-temperature phase, and gradually converging to a near-optimal configuration as the temperature cools down. The algorithm jointly optimizes the RIS location $\mathbf{l}_r$ and the initial entanglement generation rate vector $\mathbf{R}_{\text{in}}$, subject to quantum network QoS constraints, while accounting for dynamic RIS behavior (passive vs. active reflection modes) that influence channel gains and entanglement fidelity. The computational steps are detailed in Algorithm~\ref{alg:algorithm_ris}.
\begin{algorithm}[ht]
\caption{Simulated Annealing-Based Solution for $\mathcal{P}_1$}
\label{alg:algorithm_ris}
\begin{algorithmic}[1]
\State \textbf{Input:} Initial temperature $T_0$, minimum temperature $T_{\min}$, cooling rate $\alpha$, maximum iterations per temperature $L$
\State \textbf{Initialize:} Random feasible solution $(\mathbf{R}_{\text{in}}, \mathbf{l}_r, \boldsymbol{\delta})$
\State Set $T_{\text{SA}} \gets T_0$, best solution $(\mathbf{R}_{\text{in}}^*, \mathbf{l}_r^*, \boldsymbol{\delta}^*) \gets (\mathbf{R}_{\text{in}}, \mathbf{l}_r, \boldsymbol{\delta})$
\Repeat
    \For{$i = 1$ to $L$}
        \State Generate neighbor $(\mathbf{R}_{\text{in}}', \mathbf{l}_r', \boldsymbol{\delta}')$ via local perturbation
        \If{all constraints (\ref{eq:opt_obj_delta})--(\ref{eq:cons_delta}) are satisfied}
            \State Compute utility $\Delta U \gets U_{\text{WFI}}(\mathbf{l}_r', \mathbf{R}_{\text{in}}') - U_{\text{WFI}}(\mathbf{l}_r, \mathbf{R}_{\text{in}})$
            \If{$\Delta U > 0$}
                \State Accept new solution: $(\mathbf{R}_{\text{in}}, \mathbf{l}_r, \boldsymbol{\delta}) \gets (\mathbf{R}_{\text{in}}', \mathbf{l}_r', \boldsymbol{\delta}')$
            \Else
                \State Generate random $r \sim \mathcal{U}(0,1)$
                \If{$\exp(\Delta U / T_{\text{SA}}) > r$}
                    \State Accept worse solution: $(\mathbf{R}_{\text{in}}, \mathbf{l}_r, \boldsymbol{\delta}) \gets (\mathbf{R}_{\text{in}}', \mathbf{l}_r', \boldsymbol{\delta}')$
                \EndIf
            \EndIf
            \If{$U_{\text{WFI}}(\mathbf{l}_r, \mathbf{R}_{\text{in}}) > U_{\text{WFI}}(\mathbf{l}_r^*, \mathbf{R}_{\text{in}}^*)$}
                \State Update best solution: $(\mathbf{R}_{\text{in}}^*, \mathbf{l}_r^*, \boldsymbol{\delta}^*) \gets (\mathbf{R}_{\text{in}}, \mathbf{l}_r, \boldsymbol{\delta})$
            \EndIf
        \EndIf
    \EndFor
    \State Cool temperature: $T_{\text{SA}} \gets \alpha \cdot T_{\text{SA}}$
\Until{$T_{\text{SA}} < T_{\min}$}
\State \textbf{Output:} Optimal RIS location $\mathbf{l}_r^*$, EGR vector $\mathbf{R}_{\text{in}}^*$, and RIS mode configuration $\boldsymbol{\delta}^*$
\end{algorithmic}
\end{algorithm}

\section{Simulation Results and Analysis} \label{sec:simulations}
We perform extensive simulations to evaluate the performance of the proposed dynamic RIS-assisted wireless QN. For our simulations, we define the following default QN setup: A QBS is located at $\mathbf{l}_s = (0, 0, 90)\,\text{m}$ in a 3D grid, serving a number of QN users whose locations vary along the $x$- and $y$-axes, with a fixed height of 10\,m. User locations are non-uniformly distributed, following a truncated normal distribution, within a $400\,\text{m} \times 400\,\text{m}$ squared region. Each user $i$ is located at $\mathbf{l}_i = (x_i, y_i, 10)$\,m, where $x_i \sim \mathcal{T}\mathcal{N}(250, 50, 50, 450)$\,m with a mean of 250 and a standard deviation of 50, and $y_i \sim \mathcal{T}\mathcal{N}(200, 50, 0, 400)$\,m, with a mean of 200 and a standard deviation of 50, $\forall i \in \mathcal{N}$. Statistical results are averaged over 1000 simulation runs, and the region, $\Omega$, within which the dynamic RIS can be deployed corresponds to the same squared user area, defined as $\Omega = \{(x_r, y_r, H_r)\ |\ 50\,\text{m} \leq x_r \leq 450\,\text{m},\ 0\,\text{m} \leq y_r \leq 400\,\text{m},\ 35\,\text{m} \leq H_r \leq 90\,\text{m}\}$. The RIS placement location $\mathbf{l}_r$ is optimized for each set of user locations, with a minimum distance of 20\,m required between the RIS and each user to avoid being co-located. 

To represent the THz wireless medium, our simulations incorporate key THz-specific propagation effects such as molecular absorption loss, directional beam spreading, and severe path loss characteristics. The atmospheric attenuation is modeled using frequency-dependent absorption coefficients $\kappa(f)$, which become prominent in the THz range due to water vapor and oxygen resonance. A typical THz absorption coefficient of $\kappa = 0.43,\text{dB/km}$ is assumed at 300,GHz~\cite{basar2019wireless}, and beam divergence is considered using Gaussian beam parameters to reflect realistic beam shaping between the QBS-RIS-user chain. Moreover, pointing errors and turbulence—already critical in free-space optical systems—are accentuated in the THz regime due to tighter beamwidths, necessitating precise alignment via RIS control. These characteristics directly influence the photon success probability and channel transmissivity used in entanglement fidelity calculations, making dynamic RIS configuration essential to preserve quantum coherence and maximize performance under THz propagation impairments.

The feasible range of values for the initial EGR, $R_{\text{in},i}$, for a user $i \in \mathcal{N}$ is: 1\,kHz–1\,MHz~\cite{qhammami2022rate, yang2022ris}. Here, we adopt the $\alpha_s$-model for entanglement generation~\cite{chehimi2023quantumMetaverse, chehimi2024foundations}. Then, a twirling map is assumed to be applied to transform the generated entangled pairs at the QBS into identical Bell-diagonal quantum states. In particular, the initial entanglement generation attempt rate is considered to be 1\,MHz~\cite{yang2022ris}, with a probability of success for the initial entangled pair generation process $p_{\text{in}} = 2\alpha_s$, where $\alpha_s$ is a tunable parameter that enables the rate-fidelity tradeoff~\cite{chehimi2023quantumMetaverse}. We thus set the initial EGR, $R_{\text{in},i} = 2\alpha_s \times 10^6$. 

Note that the parameter $\alpha_s \in (0, 0.5)$ is directly related to the initial fidelity of the generated entangled pairs, $\lambda_{00,i}$, since $\lambda_{00,i} = 1 - \alpha_s$. As such, for an initial fidelity in the practical range of $\lambda_{00,i} \in (0.5, 0.9995)$, the range of $\alpha_s$ is adjusted to $\alpha_s \in (0.0005, 0.5)$. In particular, a very high initial EGR $R_{\text{in},i}$ corresponds to entangled qubits with a low initial fidelity $\lambda_{00,i} = 0.5$, while a low initial EGR $R_{\text{in},i}$ results in entangled qubits with a high initial fidelity $\lambda_{00,i} = 0.9995$. This corresponds to the rate-fidelity tradeoff within quantum repeater-assisted entanglement schemes, whereby the fidelity of generated entangled qubits is compromised when a high EGR is requested.
Unless stated otherwise, the minimum required fairness $\delta_{\text{th}} = 0.95$, and users' minimum E2E fidelity is sampled from a uniform distribution such that $f_{\text{min},i} \sim \mathcal{U}(0.8, 0.95)$~\cite{ chehimi2023quantumMetaverse}, which corresponds to different quantum applications. The different environmental parameters considered in the default setup, which correspond to sunny weather, moderate turbulence, and low pointing error, are summarized in Table~\ref{tab:parameters}, based on~\cite{liao2017long, basar2019wireless, yang2022ris, chehimi2023quantumMetaverse, pirandola2021limits}. Unless stated otherwise, the default setup parameters are adopted throughout the simulations.
The simulation parameters are summarized in Table~\ref{tab:parameters}, based on values from~\cite{liao2017long, basar2019wireless, chehimi2023quantumMetaverse, pirandola2021limits}.
\begin{table}[h]
\caption{Summary of Parameters Used in Simulations}
\centering
\begin{tabular}{|c|c||c|c|}
\hline
\textbf{Symbol} & \textbf{Value} & \textbf{Symbol} & \textbf{Value} \\ \hline
$\lambda$ & 1550\,nm & $\zeta_{\text{th}}$ & 0.05 \\ \hline
$\eta$ & 0.95 & $\zeta$ & 0.97 \\ \hline
$\kappa$ & 0.43\,dB/km & $C_n^2$ & \begin{tabular}[c]{@{}c@{}}moderate: $5 \times 10^{-14}$\\ strong: $1 \times 10^{-13}$\end{tabular} \\ \hline
$r_a$ & 0.55\,m & $\phi_d$ & 8\,mrad \\ \hline
$\sigma_\theta$ & 1\,mrad & $\sigma_\phi$ & 0.25\,mrad \\ \hline
$C_{\max}$ & $1 \times 10^7$ & $T$ & 2.43\,ms \\ \hline
$t_{\text{proc}}$ & 1\,$\mu$s & $\alpha_s$ & (0.0005, 0.5) \\ \hline
$\lambda_{00}$ & (0.5, 0.9995) & $R_{\text{in}}$ & 1\,kHz – 1\,MHz \\ \hline
$\Omega$ & \begin{tabular}[c]{@{}c@{}}$x_r \in [50, 450]$\,m\\ $y_r \in [0, 400]$\,m\\ $H_r \in [35, 90]$\,m\end{tabular} & $\delta_{\text{th}}$ & 0.95 \\ \hline
$f_{\min}$ & $\sim \mathcal{U}(0.8, 0.95)$ & $P_{\text{succ}}$ & Varies with $\mathbf{l}_r$ \\ \hline
\end{tabular}
\label{tab:parameters}
\end{table}

\begin{figure}[ht]
\centering
\includegraphics[width=0.65\linewidth]{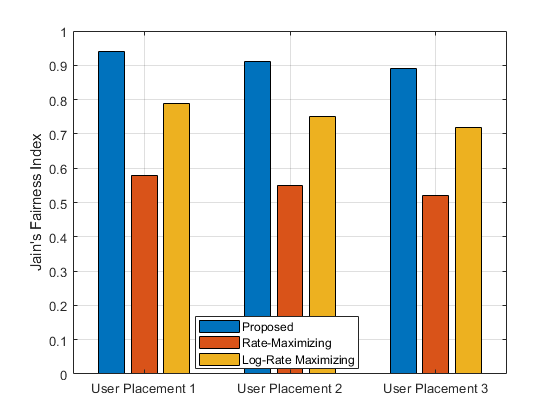}
\caption{Jain’s fairness index comparison among proposed, rate-maximizing, and log-rate maximizing schemes under three user placement scenarios.}
\label{fig:fairness}
\end{figure}

\begin{figure}[ht]
\centering
\includegraphics[width=0.65\linewidth]{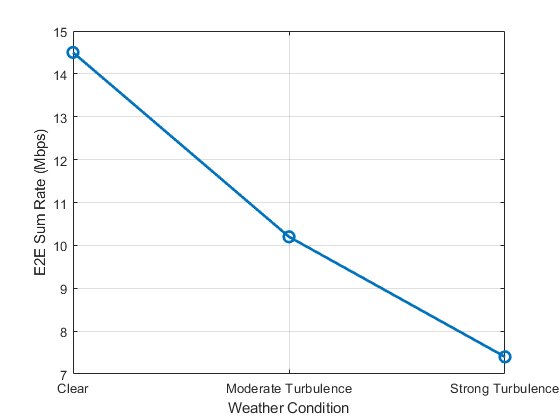}
\caption{Impact of environmental turbulence on the total end-to-end sum rate for the dynamic RIS-assisted quantum network.}
\label{fig:turbulence}
\end{figure}

In Fig.~\ref{fig:fairness}, we compare the fairness performance of three RIS-assisted quantum networking schemes under three different user placement scenarios. The proposed dynamic RIS-based framework significantly outperforms both the rate-maximizing and log-rate maximizing schemes in terms of Jain's fairness index, with improvements exceeding 60\% in certain cases. This validates the effectiveness of fairness-aware optimization that jointly considers RIS control and entanglement rate allocation.

Fig.~\ref{fig:turbulence} illustrates the sensitivity of the proposed approach to environmental turbulence conditions. As turbulence strength increases from clear to strong, the end-to-end sum rate naturally degrades due to higher channel variability. However, the system maintains reliable performance even under harsh conditions, demonstrating the robustness of the proposed RIS placement and rate control strategy in dynamically adapting to degrading channel environments.

\begin{figure*}[ht]
\label{fig:dynamic_ris_results}
    \centering  
    \begin{subfigure}[b]{0.24\linewidth}
        \centering
        \includegraphics[width=\linewidth]{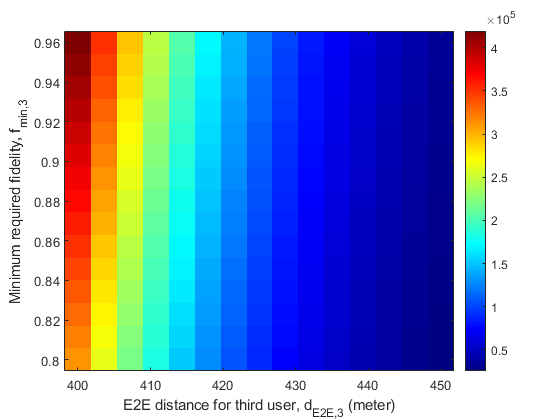}
        \caption{Moderate turbulence.}
        \label{fig:e2e_moderate}
    \end{subfigure} 
    \vspace{1em}
    \begin{subfigure}[b]{0.24\linewidth}
        \centering
        \includegraphics[width=\linewidth]{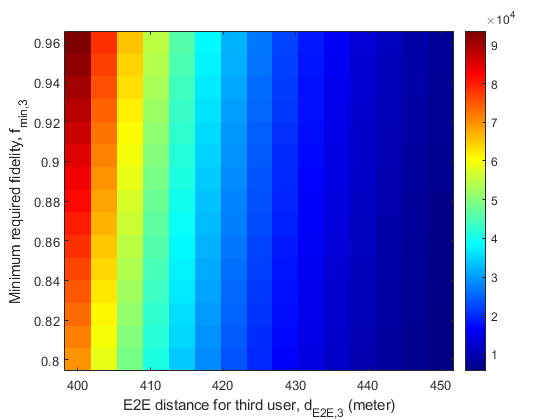}
        \caption{Strong turbulence.}
        \label{fig:e2e_strong}
    \end{subfigure}   
        \begin{subfigure}[b]{0.24\linewidth}
        \centering
        \includegraphics[width=\linewidth]{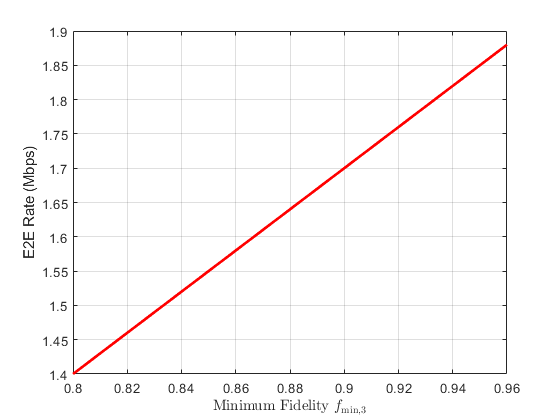}
        \caption{E2E rate versus fidelity.}
        \label{fig:e2e_Photon}
    \end{subfigure}  
        \begin{subfigure}[b]{0.24\linewidth}
        \centering
        \includegraphics[width=\linewidth]{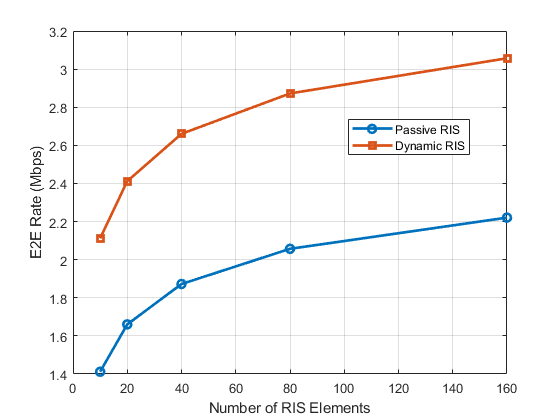}
        \caption{Effect of RIS element count}
        \label{fig:QuantumAwareTHzCapacityvsDistance}
    \end{subfigure}    
    \begin{subfigure}[b]{0.22\linewidth}
        \includegraphics[width=\linewidth]{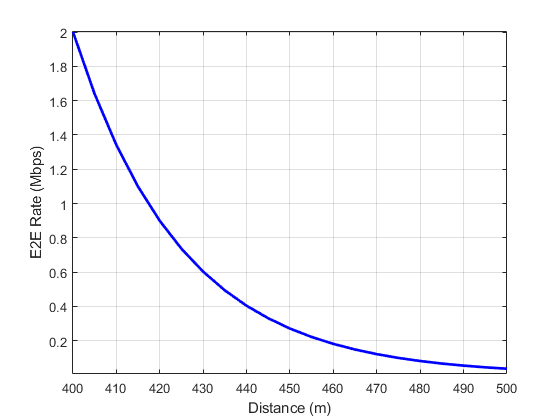}
        \caption{E2E rate as a function of communication distance.}
        \label{fig:rate_distance}
    \end{subfigure}
    \hfill
    \begin{subfigure}[b]{0.22\linewidth}
            \includegraphics[width=\linewidth]{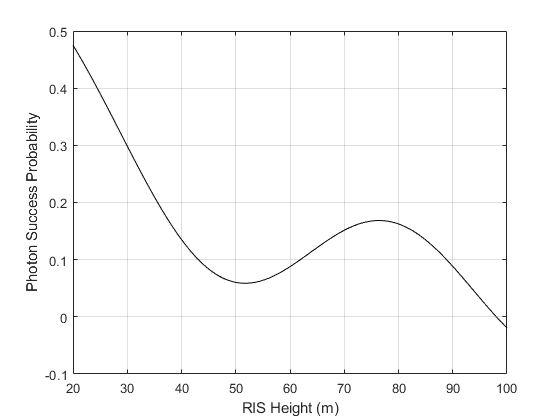}
        \caption{Photon Success vs RIS Height under THz Fading}
        \label{fig:rate_fidelity}
    \end{subfigure} 
    \vspace{1em}
    \begin{subfigure}[b]{0.22\linewidth}
        \includegraphics[width=\linewidth]{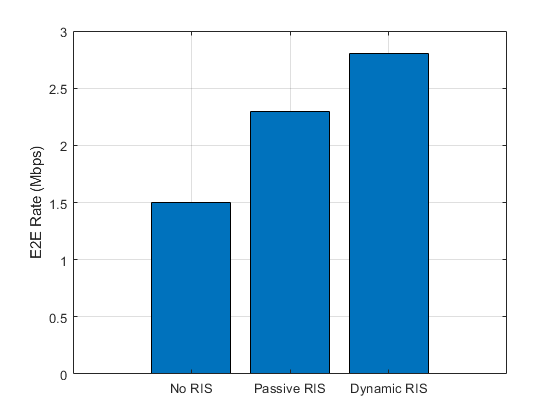}
        \caption{Comparison between no, passive, and dynamic RIS.}
        \label{fig:ris_comparison}
    \end{subfigure}
    \hfill
    \begin{subfigure}[b]{0.22\linewidth}
    \includegraphics[width=\linewidth]{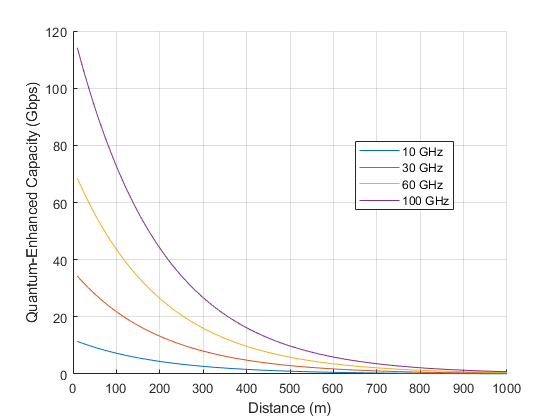}
        \caption{Quantum-Aware THz Capacity vs Distance}
        \label{fig:ris_elements}
    \end{subfigure}  
    \caption{Simulation results illustrating the impact of turbulence, THz fading, and dynamic RIS on E2E quantum network performance. (a)--(b): Rate loss under moderate/strong turbulence. (c): Tradeoff between fidelity and rate. (d): RIS element scaling benefits. (e): Distance-induced rate decay under THz. (f): Photon success vs. RIS height. (g): RIS mode comparison. (h)Quantum-aware THz capacity declines with range.}
\end{figure*}

Fig.~5 presents four detailed evaluations that highlight the impact of THz-specific phenomena on the performance of dynamic RIS-assisted quantum networks. Fig.~\ref{fig:e2e_moderate} and~\ref{fig:e2e_strong} demonstrate how E2E entanglement rate for a selected user varies under moderate and strong atmospheric turbulence conditions, respectively—emphasizing the performance drop as fidelity requirements and E2E distance increase. In Fig.~\ref{fig:e2e_Photon}, we assess the photon success probability as a function of RIS height under THz fading, revealing critical geometric alignment effects that influence quantum link reliability. Finally, Fig.~\ref{fig:QuantumAwareTHzCapacityvsDistance} illustrates the quantum-enhanced capacity of THz channels as a function of distance across different bandwidths, showing that although THz capacity rapidly decays with distance, higher-frequency bands can still maintain robust performance when coupled with quantum encoding and dynamic RIS control.

The simulation results depicted in Fig.~5 offer a comprehensive performance evaluation of the proposed quantum THz network with dynamic RIS support. Fig.~\ref{fig:e2e_moderate} and~\ref{fig:e2e_strong} show the achieved E2E entanglement rate for the third user in a 3-user QN under moderate and strong atmospheric turbulence, respectively. As expected, the E2E rate deteriorates with stronger turbulence, especially at longer distances or higher fidelity requirements. In Fig.~\ref{fig:e2e_Photon}, we examine how increasing the minimum fidelity $f_{\min}$ constrains the E2E rate due to reduced tolerance for quantum noise and decoherence. The impact of RIS size is analyzed in Fig.~\ref{fig:QuantumAwareTHzCapacityvsDistance}, where increasing the number of RIS elements—particularly in dynamic mode—substantially enhances the E2E throughput.
Fig.~\ref{fig:rate_distance} through Fig.~\ref{fig:ris_elements} delve deeper into THz-specific quantum networking factors. In Fig.~\ref{fig:rate_distance}, the E2E rate sharply decays with distance, highlighting the influence of THz attenuation and the importance of optimal link geometry. Fig.~\ref{fig:rate_fidelity} captures photon success probability as a function of RIS height under THz fading conditions, revealing non-trivial dependencies due to beam divergence and alignment uncertainty. Fig.~\ref{fig:ris_comparison} compares configurations with no RIS, passive RIS, and dynamic RIS, demonstrating how programmable RIS configurations substantially outperform static or absent RIS scenarios. Finally, Fig.~\ref{fig:ris_elements} shows the quantum-aware capacity versus communication distance across different THz bands (10–110 GHz), indicating that although higher-frequency bands offer more capacity, they also suffer faster decay—necessitating careful frequency and RIS placement strategies.
\begin{figure*}[]
    \centering   
    \begin{subfigure}[b]{0.22\linewidth}
        \includegraphics[width=\linewidth]{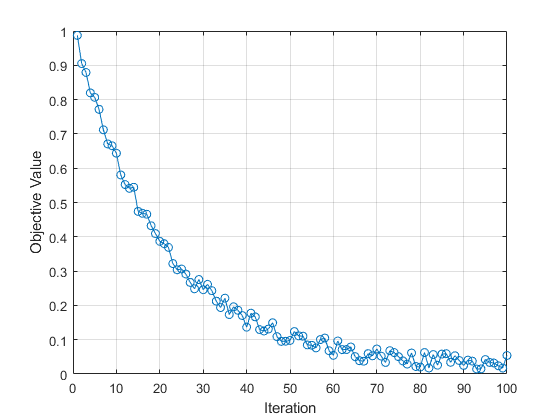}
        \caption{Simulated Annealing Convergence}
        \label{fig:Annealing}
    \end{subfigure}
    \hfill
    \begin{subfigure}[b]{0.22\linewidth}
        \includegraphics[width=\linewidth]{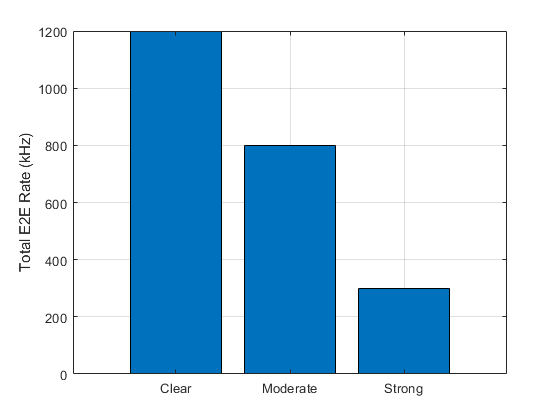}
        \caption{Turbulence Impact on E2E Sum Rate}
        \label{fig:Turbulence}
    \end{subfigure} 
    \vspace{1em}
    \begin{subfigure}[b]{0.22\linewidth}
        \includegraphics[width=\linewidth]{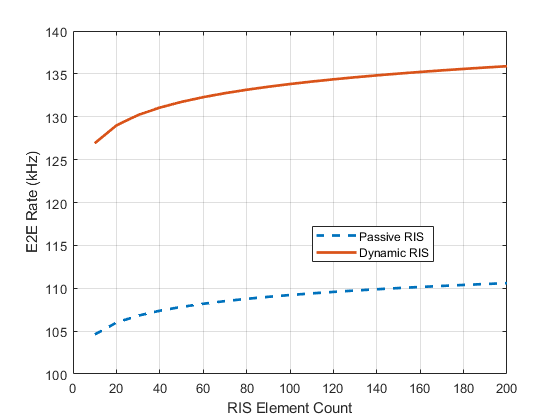}
        \caption{Impact of RIS Element Count }
        \label{fig:RISElementCount}
    \end{subfigure}
    \hfill
    \begin{subfigure}[b]{0.22\linewidth}
        \includegraphics[width=\linewidth]{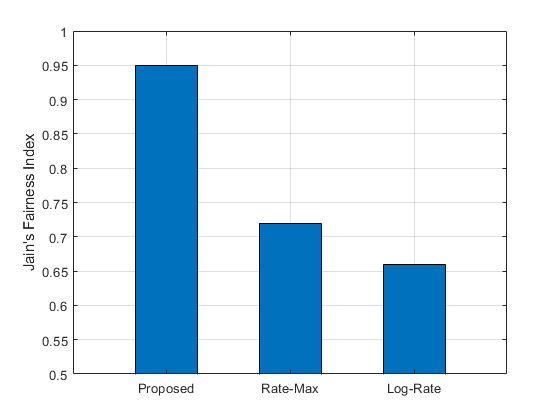}
        \caption{Fairness Comparison Among Schemes}
        \label{fig:FairnessComparisonAmongSchemes}
    \end{subfigure}  
        \hfill
    \begin{subfigure}[b]{0.22\linewidth}
        \includegraphics[width=\linewidth]{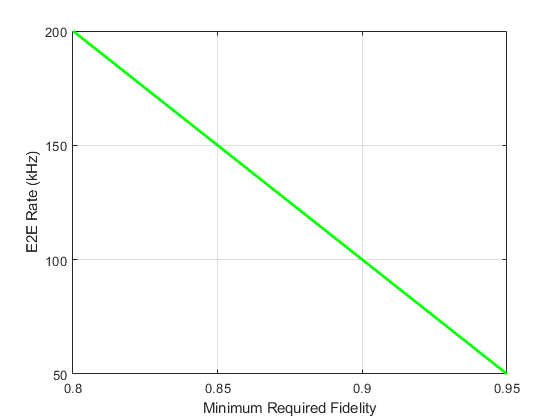}
        \caption{E2E Rate vs. Minimum Required Fidelity}
        \label{fig:E2ERatevsMinimumRequiredFidelity}
    \end{subfigure} 
        \hfill
    \begin{subfigure}[b]{0.22\linewidth}
        \includegraphics[width=\linewidth]{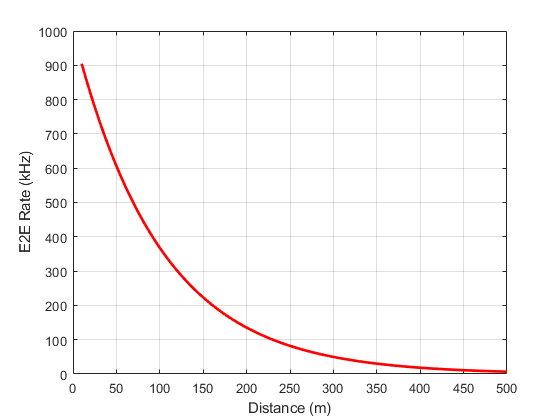}
        \caption{E2E Rate vs. Communication Distance}
        \label{fig:E2ERate}
    \end{subfigure} 
            \hfill
    \begin{subfigure}[b]{0.22\linewidth}
        \includegraphics[width=\linewidth]{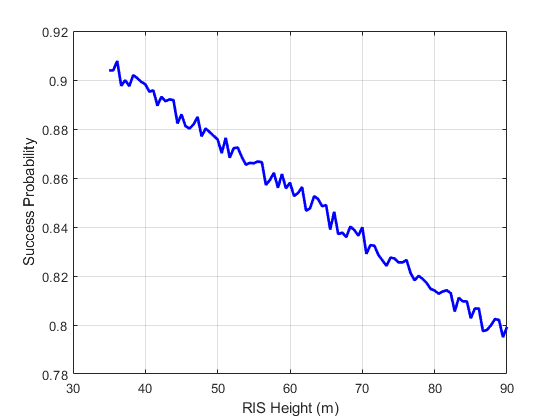}
        \caption{RIS Height vs. Photon Success Probability}
        \label{fig:Photon}
    \end{subfigure}
        \begin{subfigure}[b]{0.22\linewidth}
        \includegraphics[width=\linewidth]{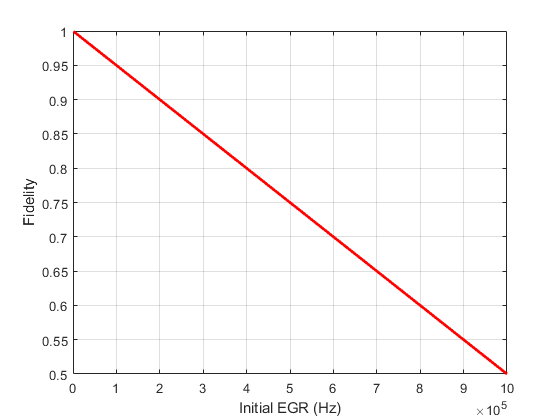}
        \caption{Rate-Fidelity Tradeoff Between EGR vs Fidelity }
        \label{fig:RateFidelityTradeoff}
    \end{subfigure}
    \caption{Illustrative results for the dynamic RIS-assisted THz quantum network. (a) Convergence of annealing-based RIS/EGR optimization. (b) E2E rate drop with turbulence. (c) Rate gains with RIS element scaling. (d) Fairness across allocation schemes. (e) Rate vs. fidelity. (f) Rate vs. distance. (g) Photon success vs. RIS height. (h) $\alpha_s$-driven EGR–fidelity tradeoff.}
    \label{fig:dynamic_ris_QN}
\end{figure*}

\section{Conclusion} \label{sec:conclusion}
In this paper, we proposed the use of an RIS in a wireless star-shaped QN that is replete with blockages under various environmental conditions. We analyzed the various losses experienced by the quantum signals over the wireless channel and proposed a novel model for the resulting quantum noise and its effect on the quality of quantum signals. We developed a novel framework that jointly optimizes the RIS placement and the initial EGR allocation among the different QN users. This joint optimization was solved while ensuring a minimum fairness level between the users and satisfying their heterogeneous QoS requirements on minimum rate and fidelity levels. Simulation results showed that our framework outperforms existing classical resource allocation frameworks that also fail to satisfy the minimum fidelity requirements. Additionally, we have verified the scalability of our framework, showed its greater sensitivity to weather conditions over pointing errors and turbulence, and found that a user's E2E distance more strongly influences the E2E rate than minimum required fidelity does.
Important future research directions include the analysis of continuous variable quantum communication sources instead of the considered discrete variable, single-photon sources. 
\appendix
\section*{Appendix A\\Proof of Theorem 1}
To prove Theorem~1, we derive the success probability for the flying qubit's arrival at user $i \in \mathcal{N}$, accounting for key impairments in the THz quantum channel: path loss, small-scale fading, and pointing misalignment. Specifically, we characterize the composite gain $h_i$, and derive the probability that this gain exceeds the quantum usability threshold $\zeta_{\text{th}}$, required to preserve entanglement fidelity.
The composite channel gain is expressed as:
\begin{equation}
h_i = h_i^{\text{pl}} \cdot h_i^{\text{a}} \cdot h_i^{\text{p}},
\end{equation}
where $h_i^{\text{pl}}$ denotes deterministic path loss, $h_i^{\text{a}}$ is a Rician fading random variable, and $h_i^{\text{p}}$ follows a bounded power-law distribution due to pointing error.
We aim to derive the PDF of $h_i$ via marginalization:
\begin{equation}
f_{h_i}(h_i) = \int f_{h_i | h_i^{\text{a}}}(h_i | h_i^{\text{a}}) f_{h_i^{\text{a}}}(h_i^{\text{a}}) \, dh_i^{\text{a}}, \label{eq:pdf-combined}
\end{equation}
where the conditional PDF given $h_i^{\text{a}}$ is:
\begin{equation}
f_{h_i | h_i^{\text{a}}}(h_i | h_i^{\text{a}}) = \frac{\vartheta_i}{A_{0,i}^{\vartheta_i} h_i^{\text{a}} h_i^{\text{pl}}} 
\left( \frac{h_i}{h_i^{\text{a}} h_i^{\text{pl}}} \right)^{\vartheta_i - 1},
\end{equation}
valid for $0 \le h_i \le A_{0,i} h_i^{\text{a}} h_i^{\text{pl}}$.
The small-scale fading component $h_i^{\text{a}}$ is modeled using a Rician distribution, whose PDF involves the modified Bessel function of the second kind, $K_{\nu}(\cdot)$. Substituting this into Eq.~\eqref{eq:pdf-combined}, we obtain:
\begin{align}
&f_{h_i}(h_i) = \frac{2 \vartheta_i (\alpha_i \beta_i)^{(\alpha_i + \beta_i)/2}}{(A_{0,i} h_i^{\text{pl}})^{\vartheta_i} \Gamma(\alpha_i)\Gamma(\beta_i)} 
h_i^{\vartheta_i - 1} \times
\nonumber \\
& \int_{h_i / A_{0,i} h_i^{\text{pl}}}^{\infty}
(h_i^{\text{a}})^{\frac{\alpha_i + \beta_i}{2} - 1 - \vartheta_i}
K_{\alpha_i - \beta_i} \left( 2 \sqrt{\alpha_i \beta_i h_i^{\text{a}}} \right) dh_i^{\text{a}}. \label{eq:combined-int}
\end{align}

To express this integral in closed form, we employ the identity for the Bessel function in terms of the Meijer G-function:
\begin{equation}
K_{\alpha - \beta}(x) = \frac{1}{2} \mathcal{G}_{0,2}^{2,0}\left[ \frac{x^2}{4} \left| 
\begin{array}{c}
- \\
\frac{\alpha - \beta}{2}, \frac{\beta - \alpha}{2}
\end{array}
\right. \right]. \label{eq:bessel-g}
\end{equation}

Using convolution and transformation rules for G-functions, the final PDF becomes:
\begin{align}
f_{h_i}(h_i) =& \frac{\vartheta_i \alpha_i \beta_i}{A_{0,i} h_i^{\text{pl}} \Gamma(\alpha_i)\Gamma(\beta_i)} 
\mathcal{G}_{1,3}^{3,0} \nonumber \\
& \left[ 
\left. \frac{\alpha_i \beta_i}{A_{0,i} h_i^{\text{pl}}} \,\right| 
\begin{array}{c}
\vartheta_i \\
\vartheta_i - 1, \alpha_i - 1, \beta_i - 1
\end{array}
\right]. \label{eq:simplified-fhi}
\end{align}

Now, we define the probability of successful photon transmission as:
\begin{equation}
P_{\text{succ}}(\zeta_{\text{th}}) = \mathbb{P}(h_i > \zeta_{\text{th}}) 
= 1 - \int_0^{\zeta_{\text{th}}} f_{h_i}(h_i) dh_i. \label{eq:psucc2}
\end{equation}

This integral can be resolved via standard G-function integration identities, yielding the final expression:
\begin{align}
&P_{\text{succ},i}(\mathbf{l}_r) = 1 
- \left( \frac{\vartheta_i(\mathbf{l}_r)}{\Gamma(\alpha_i(\mathbf{l}_r)) \Gamma(\beta_i(\mathbf{l}_r))} \right) 
\times \nonumber \\
&\quad  \mathcal{G}_{2,4}^{3,1} \left[ 
\left. \frac{\alpha_i(\mathbf{l}_r) \beta_i(\mathbf{l}_r) \chi_{\text{th}}}
{A_{0,i}(\mathbf{l}_r) h_i^{\text{pl}}(\mathbf{l}_r)} \,\right| 
\begin{array}{l}
1,\ \vartheta_i(\mathbf{l}_r)+1 \\
\vartheta_i(\mathbf{l}_r),\ \alpha_i(\mathbf{l}_r),\ \beta_i(\mathbf{l}_r),\ 0
\end{array}
\right], \label{eq:final-proof}
\end{align}
where $\chi_{\text{th}} = \frac{\zeta_{\text{th}}}{\eta}$ captures normalized fidelity threshold. This completes the proof. \hfill

\section*{Appendix B\\Proof of Proposition 1}
We now derive the closed-form expression for the E2E entanglement fidelity between the QBS and user $i \in \mathcal{N}$, accounting for the composite quantum noise introduced by the memory decoherence and THz RIS-assisted wireless quantum transmission.
We begin by analyzing the decoherence on the matter qubit stored at the QBS. The noise is modeled as a phase damping channel $\Lambda_{p_1}$ acting on the first qubit of the Bell-diagonal state $\rho_{\text{BD},i}$:
\begin{equation}
\Lambda_{p_1}(\rho) = (1 - p_1)\rho + p_1 \sigma_z \rho \sigma_z,
\end{equation}
where $p_1$ is the memory-induced phase damping probability. Applying this map to the first qubit of $\rho_{\text{BD},i}$ results in:
\begin{equation}
\rho_{\text{BD},i}^{'} = (1 - p_1) \rho_{\text{BD},i} + p_1 (\sigma_z \otimes \mathbb{I}) \rho_{\text{BD},i} (\sigma_z \otimes \mathbb{I}).
\end{equation}

The Pauli-Z operator flips the phase of $|1\rangle$, and when acting on Bell states, results in:
\[
\begin{aligned}
(\sigma_z \otimes \mathbb{I})\Phi_{00} &= \Phi_{01}, & \quad (\sigma_z \otimes \mathbb{I})\Phi_{01} &= \Phi_{00}, \\
(\sigma_z \otimes \mathbb{I})\Phi_{10} &= \Phi_{11}, & \quad (\sigma_z \otimes \mathbb{I})\Phi_{11} &= \Phi_{10}.
\end{aligned}
\]

Hence, the resulting state becomes:
\begin{align}
\rho_{\text{BD},i}^{'} &= (1 - p_1)(\lambda_{00,i}\Phi_{00} + \lambda_{01,i}\Phi_{01} + \lambda_{10,i}\Phi_{10} + \lambda_{11,i}\Phi_{11}) \nonumber \\
&\quad + p_1(\lambda_{00,i}\Phi_{01} + \lambda_{01,i}\Phi_{00} + \lambda_{10,i}\Phi_{11} + \lambda_{11,i}\Phi_{10}) \nonumber \\
&= \sum_{j,k \in \{0,1\}} V_{jk,i} \Phi_{jk}, \label{eq:p1_transform}
\end{align}
where
\begin{equation}
V_{jk,i} = (1 - p_1)\lambda_{jk,i} + p_1 \lambda_{j(k \oplus 1),i}.
\end{equation}

Next, the flying qubit experiences a phase damping channel during RIS-assisted THz transmission. Applying this to the second qubit yields:
\begin{equation}
\rho_{\text{BD},i}^{''}(\mathbf{l}_r) = (1 - P_{\text{pd},i}^{(2)}(\mathbf{l}_r))\rho_{\text{BD},i}^{'} + P_{\text{pd},i}^{(2)}(\mathbf{l}_r)(\mathbb{I} \otimes \sigma_z)\rho_{\text{BD},i}^{'}(\mathbb{I} \otimes \sigma_z).
\end{equation}

Again using Bell state identities:
\[
\begin{aligned}
(\mathbb{I} \otimes \sigma_z)\Phi_{00} &= \Phi_{01}, & \quad (\mathbb{I} \otimes \sigma_z)\Phi_{01} &= \Phi_{00}, \\
(\mathbb{I} \otimes \sigma_z)\Phi_{10} &= \Phi_{11}, & \quad (\mathbb{I} \otimes \sigma_z)\Phi_{11} &= \Phi_{10},
\end{aligned}
\]
where
\begin{equation}
\rho_{\text{BD},i}^{''}(\mathbf{l}_r) = \sum_{j,k} U_{jk,i} \Phi_{jk},
\end{equation}
\begin{equation}
U_{jk,i}(\mathbf{l}_r) = (1 - P_{\text{pd},i}^{(2)}(\mathbf{l}_r))V_{jk,i} + P_{\text{pd},i}^{(2)}(\mathbf{l}_r)V_{j(k \oplus 1),i}.
\end{equation}

Finally, we apply amplitude damping noise on the second qubit (flying qubit) with the Kraus operators:
\begin{align}
E_{0,i} &= \mathbb{I} \otimes \left( |0\rangle \langle 0| + \sqrt{1 - P_{\text{ad},i}^{(2)}(\mathbf{l}_r)} |1\rangle \langle 1| \right), \\
E_{1,i} &= \mathbb{I} \otimes \left( \sqrt{P_{\text{ad},i}^{(2)}(\mathbf{l}_r)} |0\rangle \langle 1| \right).
\end{align}

This results in the final E2E entangled state $\rho_{\text{E2E},i}(\mathbf{l}_r)$, where non-diagonal coherence terms and population leakage terms are scaled based on $P_{\text{ad},i}^{(2)} = 1 - P_{\text{succ},i}$. The fidelity with respect to the target state $\Phi_{00}$ is then:
\begin{align}
f_{\text{E2E},i}(\mathbf{l}_r) &= \langle \Phi_{00} | \rho_{\text{E2E},i}(\mathbf{l}_r) | \Phi_{00} \rangle \nonumber \\
&= \frac{1}{4}(U_{00,i} + U_{01,i}) + \frac{1}{4}(1 - P_{\text{succ},i}(\mathbf{l}_r))(U_{10,i} + U_{11,i}) \nonumber \\
&\quad + \frac{1}{4}P_{\text{succ},i}(\mathbf{l}_r)(U_{00,i} + U_{01,i}) + \nonumber \\
& \frac{1}{2}\sqrt{P_{\text{succ},i}(\mathbf{l}_r)}(U_{00,i} - U_{01,i}). \label{eq:fidelity_final_closed}
\end{align}

This completes the proof of Proposition~1.

\bibliographystyle{IEEEtran}
\bibliography{IEEEabrv,Reference/mybib}
 
 \begin{IEEEbiography}
    [{\includegraphics[width=1in,height=1.25in,clip,keepaspectratio]{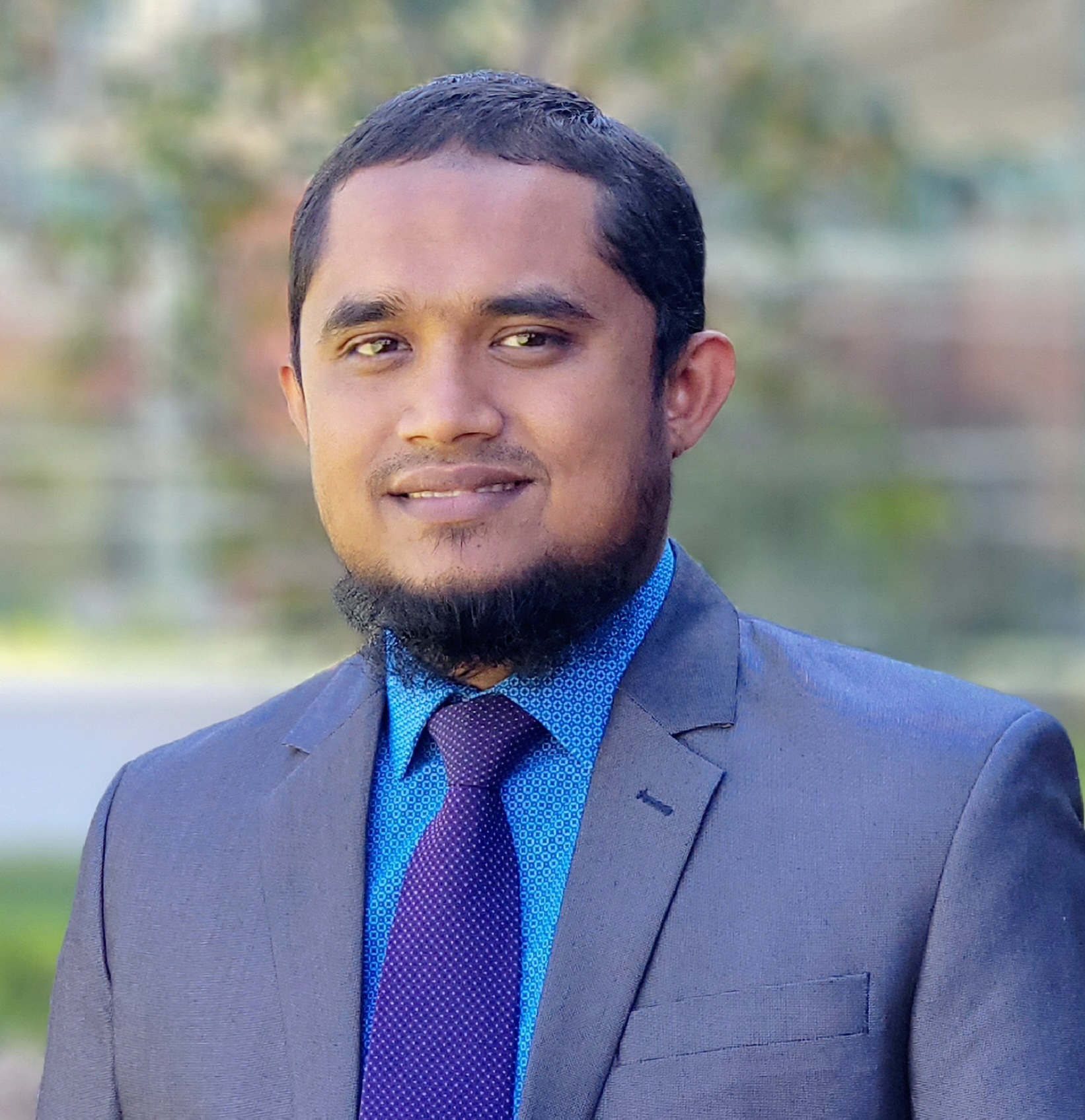}}]{Shakil Ahmed (IEEE Member)}
 is an Assistant Professor (Term) in the Department of Electrical and Computer Engineering at Iowa State University. He received his B.S. degree in Electrical and Electronic Engineering from Khulna University of Engineering \& Technology, Bangladesh, in 2014 and his M.S. degree in Electrical Engineering from Utah State University, USA, in 2019. He later earned his Ph.D. in Electrical and Computer Engineering from Iowa State University in a record time of 2 years and 8 months. Ahmed has published numerous research papers in renowned international conferences and journals and received the Best Paper Award at international venues. His research interests include cutting-edge areas such as next-generation wireless communications, GenAI, wireless network design and optimization, unmanned aerial vehicles, physical layer security, Digital Twin for wireless communications, content creation using R.F. signals, and reconfigurable intelligent systems. He is also passionate about engineering education and integrating generative A.I. into learning processes. He has also been a guest editor and reviewer for prestigious journals, including IEEE Transactions on Cognitive Communications and Networking, IEEE Access, IEEE Systems Journal, and IEEE Transactions on Vehicular Technology.
\end{IEEEbiography}

\end{document}